\documentclass[pre,twocolumn,showpacs,floatfix]{revtex4}

\usepackage{graphicx}

\begin{document}

\title{Hexagonal convection patterns in atomistically simulated fluids} 
\author{D. C. Rapaport}
\email{rapaport@mail.biu.ac.il}
\affiliation{Physics Department, Bar-Ilan University, Ramat-Gan 52900, Israel}
\date{August 07, 2005}

\begin{abstract}
Molecular dynamics simulation has been used to model pattern formation in
three-dimensional Rayleigh--B\'enard convection at the discrete-particle level.
Two examples are considered, one in which an almost perfect array of
hexagonally-shaped convection rolls appears, the other a much narrower system
that forms a set of linear rolls; both pattern types are familiar from 
experiment. The nature of the flow within the convection cells and quantitative
aspects of the development of the hexagonal planform based on automated polygon
subdivision are analyzed. Despite the microscopic scale of the system,
relatively large simulations with several million particles and integration
timesteps are involved.
\end{abstract}


\maketitle

The spontaneous emergence of flow patterns in externally driven fluids is one of
the more fascinating aspects of fluid dynamics, and a great deal of experimental
and theoretical effort has been spent studying such phenomena over the past
century. Theoretical fluid dynamics has successfully ignored the underlying
atomicity, describing the dynamics in terms of nonlinear partial differential
equations for continuous fields; analysis of flow instability within this
framework demands approximations whose consequences are not always readily
assessed. Ideally, one would like to understand flow instability in terms of the
behavior at the microscopic level by looking beyond the continuum representation
to address the dynamics of the molecules themselves, albeit at considerable
computational cost; given sufficiently many molecules followed for an adequate
time interval, one can reasonably expect to see the same kind of behavior
(initially in a qualitatively correct form, then, with even larger systems, also
quantitatively). The ability to simultaneously encompass the dynamics both at
the molecular level and the scale at which the cooperativity underlying
structured flow becomes manifest is an important step in trying to break free of
the limitations of traditional continuum fluid dynamics.

Accomplishing this goal calls for molecular dynamics (MD) simulation, a
technique in broad general use, but less so for fluid dynamics (quite likely due
to Avogadro's tyranny -- an apparent need for extremely large systems to
accommodate the phenomena). A very early MD effort related to bridging the gap
between atomistic dynamics and fluid behavior at the continuum level was
understanding long-time effects due to correlations in atomic trajectories
\cite{ald70}, but it was only much later that actual MD simulation of complex
flow was attempted. Initial efforts were confined to two dimensions, starting
with the wakes of obstructed flows \cite{rap86,rap87} and followed by
Rayleigh--B\'enard (RB) convection \cite{mar87,rap88,mar88,puh89,rap92a,rap92b};
the study of Taylor--Couette flow \cite{hir98,hir00} demonstrated that complex
three-dimensional flow could also be dealt with and, subsequently, the
Rayleigh--Taylor instability was also modeled \cite{kad04}. Other instances of
MD contributing to fluid dynamics include the moving interface between
immiscible liquids undergoing Poiseuille flow \cite{kop89} and droplet breakup
in liquid jets \cite{mos00}. The present study describes the application of the
MD approach to the full three-dimensional RB problem, where the behavior is
potentially far richer than in two dimensions, and where simulation is more
closely related to experiment; the ability to tackle increasingly demanding MD
simulations, both in terms of size and duration, reflects the availability of
increasingly powerful and affordable computing resources.

The RB phenomenon -- the rich variety of flow patterns produced by convection in
a fluid layer heated from below -- continues to be a problem of great interest
\cite{cha61,nor77,ber84,kos93,bod00}. The principal dimensionless quantity
governing the behavior is the Rayleigh number, $Ra = \alpha g L_z^3 \Delta T /
\nu \kappa$, where $\alpha$, $\nu$ and $\kappa$ are the thermal expansion
coefficient, kinematic viscosity and thermal diffusivity, $L_z$ is the layer
height, $g$ the gravitational acceleration, and $\Delta T = T_1 - T_0$ is the
temperature difference between the lower hot wall $T_1$ and the upper cold wall
$T_0$; a critical value $Ra_c$ marks the onset of convection, where  buoyancy
overcomes viscous drag, and convection replaces conduction as the preferred
thermal transport mechanism.

Theory is simplified by the Boussinesq approximation, in which it is assumed the
only temperature-dependent fluid property is the density. $Ra_c$ can be computed
for various kinds of boundary conditions \cite{cha61}, as can the wavelength
$\lambda$ of the convection pattern at criticality, but not the actual planform
(e.g., rolls, spirals and hexagons). Roll width (typically $\approx L_z$)
represents a compromise; wide rolls reduce both viscous drag and diffusive heat
transfer between ascending hot and descending cold streams, whereas narrow rolls
reduce shear losses close to nonslip walls. Hexagonal patterns were once
associated with non-Boussinesq fluids \cite{bod91,kos93} (and are better known
in convection driven by surface tension); the flow direction at the cell center
is determined by the temperature dependence of $\nu$ and differs for gases and
liquids. Hexagons are now known to occur in (essentially) Boussinesq systems
above $Ra_c$ \cite{ass96,baj97}, even forming coexisting regions with opposite
flows at the cell centers, a phenomenon not attributable to variable $\nu$.

The MD simulation \cite{rap04} considers a fluid of soft-sphere atoms with a
short-ranged, repulsive interaction $u(r) = 4 \epsilon [(\sigma / r)^{12} -
(\sigma / r)^6]$, with a cutoff at $r_c = 2^{1/6} \sigma$. In the dimensionless
MD units used subsequently, $\sigma$ and $\epsilon$ determine length and energy,
while temperature is defined by setting $k_B = 1$; in the case of Argon, $\sigma
= 3.4$\,\AA, $\epsilon / k_B = 120$\,K, and the unit of time is 2.16\,ps. The
top and bottom thermal walls of the system are each formed from a layer of fixed
atoms arranged as a lattice, spaced to ensure roughness and impenetrability;
lateral boundaries are periodic. The temperature gradient is produced by
rescaling the velocities of those atoms adjacent (within a range of 1.3) to the
thermal walls; rescaling occurs every 20 timesteps to allow the effect to
propagate without unduly affecting the dynamics, and is applied to atoms moving
away from the walls. In the initial state, atoms are placed on the sites of a
regular grid, with random velocities corresponding to a uniform vertical
temperature gradient. The integration timestep is 0.004. In view of the
extensive computations required, parallel processing based on spatial
decomposition is used.

Flow analysis entails spatial coarse-graining, in which a grid subdivision is
applied to the region and the mean properties (velocity, temperature, density)
for each grid cell are evaluated for its occupant atoms. This is repeated every
20 timesteps and averages over 100 successive measurements recorded, with
typical grid sizes of 5--6 horizontally and 3 vertically ($\approx$ 30 atoms per
grid cell); this represents a compromise between capturing the fine spatial and
temporal detail of the developing convection patterns and reducing noise due to
thermal fluctuations.

\begin{figure}
\includegraphics[scale=1.4]{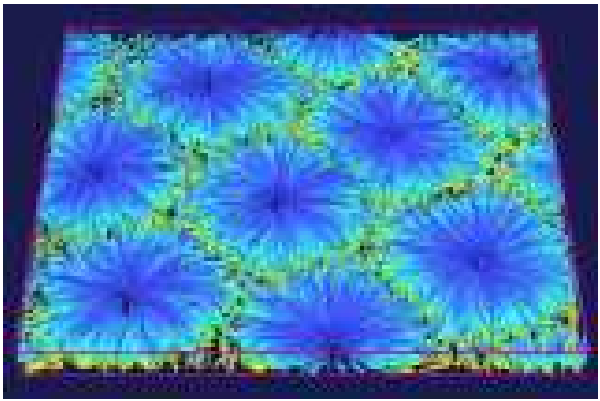}
\caption{\label{fig:1} (Color online) Final streamline plot for the large
rectangular system \#A showing the hexagonal cell pattern; the streamlines are
color coded to indicate temperature variation (ranging from red for hot to blue
for cold).}
\end{figure}

Two of a series of runs are described in detail below; the eventual outcomes of
both are organized convection cell patterns, a hexagonal array in the case of
system \#A and a set of linear rolls for \#B. System \#A contains $N_a =
3\,507\,170$ atoms (of which $N_w = 447\,458$ are fixed in the walls) and is run
for a total of $3.19 \times 10^6$ timesteps ($= 2.8 \times 10^{-8}$\,s). The
region size is $L_x \times L_y \times L_z = 521 \times 451 \times 35.3$; despite
the smallness of $L_z$ (a mere 120\,\AA), the layer thickness exceeds the value
25 used in the Taylor--Couette study \cite{hir98,hir00}. Since the rate of
pattern change slows as features become larger, a reflection of the underlying
horizontal diffusion processes, the long run is needed to ensure a final steady
state is reached. The smaller system \#B has $L_y = 65$ ($L_x$ and $L_z$ are
unchanged), resulting in a more modest $N_a = 506\,696$ (with $N_w = 64\,328$),
and a run length of $0.48 \times 10^6$ timesteps. In both cases, $T_0 = 1$, $T_1
= 10$; the value $g = 3 \Delta T / 2 L_z$ ensures equality of the potential and
kinetic energy changes across the layer and avoids excessive density variation.
Large gradients and fields are the norm for MD flow studies that must deal with
relatively small regions, here compensating for the small $L_z$ to ensure an
adequate value of $Ra$; $\Delta T / T_0$ is large compared with experiment
(where $\Delta T / T_0 \approx 0.01$ or less), and $g = 2.7 \times 10^{12} \,
g_{earth}$ (the variation in gravitational potential across the layer, $g L_z$,
is only 8 times that of \cite{kad04}). The total computation time for system \#A
amounts to approximately 6 cpu-months on a 2\,GHz Intel Pentium processor, with
the actual run using an 8-cpu cluster.

The final state of system \#A is shown in Fig.~\ref{fig:1} as a streamline plot
that reveals the full three-dimensional flow structure; the curved tracks (color
coded to show temperature variation) are derived from the coarse-grained grid
(size $96 \times 82 \times 12$). The most prominent feature is the array of
hexagonal convection cells extending across the full depth of the layer, with
cold fluid descending at the cell centers. The cell array is free to orient
itself to accommodate a preferred $\lambda$ for the particular size and shape
of the periodic container; animated three-dimensional visualization allows
following the pattern development as cells grow and merge (not shown).

\begin{figure}
\includegraphics[scale=1.19]{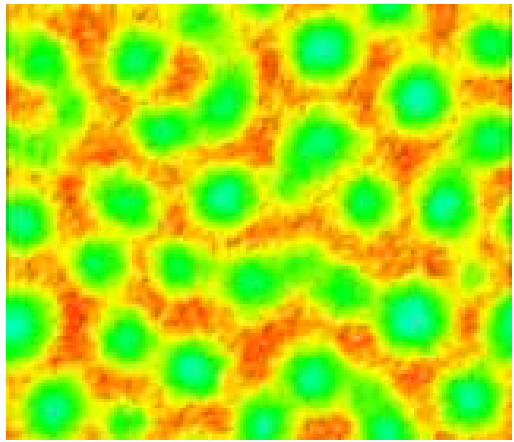}
\caption{\label{fig:2} (Color online) Early temperature distribution near the
lower wall.}
\end{figure}

\begin{figure}
\includegraphics[scale=1.19]{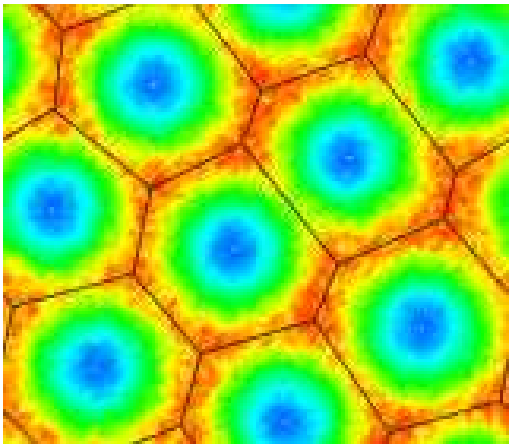}
\caption{\label{fig:3} (Color online) Final temperature distribution with
Voronoi polygon subdivision superimposed.}
\end{figure}

Figure~\ref{fig:2} shows the horizontal temperature distribution at a distance
$\approx L_z / 5$ from the hot wall early in the run, after the onset of
convection but before a regular cell pattern develops. The distribution at the
end of the run appears in Fig.~\ref{fig:3}, with the periodic cell array clearly
visible. Superimposed on this image is a polygon subdivision produced by an
automated Voronoi analysis of a slice through the flow structure (limited system
size cannot accommodate the large aspect ratios $L_x / L_z > 100$ used
experimentally, as opposed to 15 here, that yield extended patterns amenable to
Fourier and other forms of analysis \cite{ego98}); cell centers marked by white
spots are located at each of the local temperature minima (or at the downward
flow maxima) and the plane subdivided into polygons in which the region nearer
to a particular center than to any other is assigned to its polygon. The
resulting polygon set provides a good fit to the cell boundaries where hot
upward flow is strongest (the fit quality improves as the run progresses), and
is used in the quantitative analysis below. 

\begin{figure}
\includegraphics[scale=0.75]{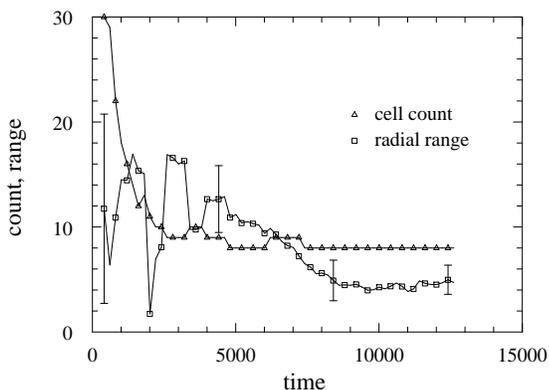}
\caption{\label{fig:4} Graphs showing time-dependence (dimensionless MD units)
of the cell count and the mean radial range (typical fluctuations are included
for the latter).}
\end{figure}

The time-dependence of two prominent features of the polygon arrays is shown in
Fig.~\ref{fig:4}. The polygon count falls from an initial value of $\approx 30$
to the eventual 8 at time $t \approx 5000$, it then increases to 9 and finally
drops back to 8 at $t \approx 8000$ as cells split and merge; this behavior is
an indication of the slow relaxation processes that the simulations must be able
to cover. The mean deviation of the polygon shape from regularity, expressed as
the range of radial distances of the nearest and furthest vertices of each
polygon, shows a more gradual convergence to the final value of 4.5, which is
just 5\% of the mean radius. Other planform properties include the mean distance
from the center to the nearest point on an edge, $\lambda_{hex} / 2$, whose
final value is $93 \pm 1$, the mean polygon area that is inversely proportional
to the count, and the area variation that falls to a minimum of $\pm$ 2\% of the
mean in the final state. The average number of vertices per polygon must be 6;
the spread of values falls to zero as the final state is neared. These results
quantify what is observed directly, namely an initially disordered set of small
convection cells that grow, merge and rearrange, culminating in what appears to
be a stable array of hexagonal cells; the final wavelength $\lambda_{hex} = 5.27
\, L_z$, a value not much larger than the experimental $4 \, L_z$ \cite{ass96}
(a lower $\lambda_{hex}$ is reported in \cite{baj97}); the characteristic size
of the cells present in the early, irregular pattern is about half this value
(Fig.~\ref{fig:2}).

\begin{figure}
\includegraphics[scale=0.5]{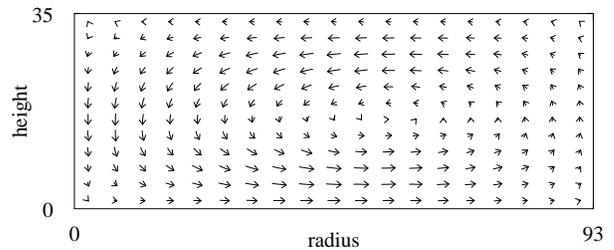}
\caption{\label{fig:5} Flow field azimuthally averaged over all cells; the axes
measure the radial distance from the cell center and the height above the hot
wall (dimensionless MD units).}
\end{figure}

Details of the flow profile inside a single convection cell appear in
Fig.~\ref{fig:5}. The arrow plot shows the radial and vertical components of the
flow field within the cells, averaged over all cells of the final state, and is
evaluated by projecting the flow velocity onto a series of vertical planes
arranged as radial spokes around each cell center, with only grid cells that
intersect the planes contributing. The maximum arrow length corresponds to a
speed of 0.52. While average descending flow appears faster than ascending, this
is consistent with the toroidal roll shape that allows more space for the
latter. The nonslip nature of the flow at the lower wall is partly masked by
coarse-graining, but flow rate is seen to decrease near the wall (the
effectiveness of the nonslip mechanism is better appreciated in small systems
run interactively). Local temperature (not shown) increases radially outwards
from the cell core and drops with height; its estimation is problematic because
the contribution of nonuniform bulk flow cannot be fully removed, and
thermalization includes atoms interacting with the walls, so the measured range
is 0.84--11.4 (rather than 1--10), but the gradual variation across the roll
profile and the absence of high gradients near the walls (except where the cool
downflow initially encounters the hot wall) are evidence of efficient heat
exchange between fluid and walls. Local density varies from 0.58 at the cell
core to 0.38 at the periphery (the overall mean is 0.4).

\begin{figure}
\includegraphics[scale=1.4]{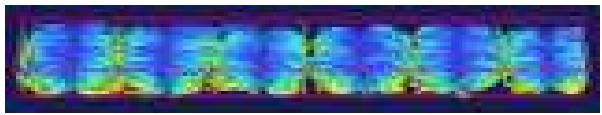}
\caption{\label{fig:6} (Color online) Final streamline plot for system \#B
showing the linear roll pattern.}
\end{figure}

The final state of run \#B consists of a set of 6 pairs of counter-rotating
rolls. Fig.~\ref{fig:6} shows the streamline plot; the wavelength
$\lambda_{roll}$ ($2 \times$ roll diameter) $= 87 = 2.46 \, L_z$ (note that 7
roll pairs would have produced a value close to the theoretical $\lambda_{roll}
= 2.016 \, L_z$ \cite{cha61}). The wavelength ratio from runs \#A and \#B is
$\lambda_{hex} / \lambda_{roll} = 2.14$; the experimental ratio for coexisting
hexagons and rolls is 1.2--1.3 \cite{ass96}. At an early stage in run \#B there
is an unsuccessful attempt to form a pair of longitudinal rolls.

When $T_1$ is lowered to 6 the roll pattern in system \#B is unchanged, but at
$T_1 = 4$ the results are noisy and roll structure is not readily discerned,
though it is still present in weakened form. Strong finite-size effects due to
small $L_z$ erase any hint of a sharp transition between conducting and
convecting states. The fluid has its own intrinsic thermodynamic and transport
properties determined by the soft-sphere model and, since $\Delta T / T_0$ is
large, does not satisfy the Boussinesq condition; estimating $Ra$ based on
Enskog theory and a simple equation of state \cite{puh89,rap92a} is
inappropriate here due to the variation of $\nu$ and $\kappa$ across the layer.

These simulations (and work in progress) provide strong confirmation that MD is
capable of reproducing known fluid dynamical behavior; the unique aspect of this
approach is that, while following the dynamics at the atomic scale where the
underlying causes of flow instability and self-organization are to be found, it
simultaneously embraces the continuum scale where fluid behavior is manifest and
scant evidence of atomism remains. Increasing computer performance (parallel
computers are ideal for simulations of this kind) allows ever-larger systems to
be modeled, although this growth is tempered by the fact that computation time
increases not only in proportion to system size but also due to slow diffusive
processes that govern structure development. Larger systems reduce the
necessarily high thermal and velocity gradients and help eliminate spurious
finite-size effects, allowing closer correspondence with experiment. The outcome
of the present MD study has obvious implications for future MD exploration of
fluid dynamical phenomena.

\begin{acknowledgments}
This work was begun during a visit to the University of Edinburgh in the last
millennium that was supported by the TRACS program at EPCC; S. Pawley is thanked
for his kind hospitality. 
\end{acknowledgments}

\bibliography{raybenmd}

\end{document}